\documentclass[12pt,preprint]{aastex}

\newcommand{\lboo}{$\lambda$~Bootis}

\newcommand{\dscuti}{$\delta$~Scuti}

\shorttitle{Mixing in \lboo\ Stars}
\shortauthors{Turcotte}

\begin{document}

\title{Mixing and Accretion in \lboo\ Stars}

\author{S. Turcotte}
\affil{Lawrence Livermore National Laboratory, L-413, P. O. Box 808,
    Livermore, CA 94551}
\email{sturcotte@igpp.ucllnl.org}

\begin{abstract}
Strong evidence for deep mixing has been uncovered for slowly rotating
F, and A stars of the main sequence. As the accretion/diffusion model
for the formation of \lboo\ stars is
heavily dependent on mixing in superficial regions, such deep mixing may
have important repercussions on our understanding of these stars. It is shown that deep mixing
at a level similar to that of FmAm stars increases the amount of
matter that needs to be accreted by the stars with respect with the
standard models by some three orders of magnitude. 
It is also shown that significantly larger accretion rates
have to be maintained, as high as $10^{-11}$~M$_\sun\,yr^{-1}$,
to prevent meridional circulation from canceling
the effect of accretion.  The existence of old ($\approx 1$~Gyr) is not 
a likely outcome of the present models for accretion/diffusion with or without
deep mixing. It is argued that \lboo\ stars are potentially very 
good diagnostics of mixing mechanisms in moderately fast rotators.
\end{abstract}

\keywords{stars: abundances --- stars: evolution --- stars: interior}

\section{Introduction}

The defining characteristic of \lboo\ stars is their peculiar surface
composition in which only four elements are roughly solar (C, N, O and S) and all others show
a definite trend toward depletion by a factor of up to ten typically 
\citep{Heiter02,Solanoetal01}. 
Yet, despite their apparent metal-poor composition these are population~I stars
in which the peculiar chemical composition is only superficial.
There are concerns that the class, as it is now defined based on their
chemical composition, does not
represent an homogeneous population resulting from a common physical 
process.  For example, 
it is now clear that there is a wide scatter in the abundances of
individual elements from star to star, larger than in chemically normal
A-type stars \citep{Heiter02}. 
Other potentially
conflicting observations and the failure of models to reproduce all 
the characteristics of \lboo\ stars, which will be discussed briefly
in this paper, caution us to keep that possibility in mind.
 
In addition to their chemical signature, this class of stars is also
characterized by a limited range in spectral types (early F and A type
stars) \citep{Solanoetal01}.
For a long time they were also thought to be strictly limited to young
stars (ZAMS or pre-main-sequence) but the evidence now points to
ages ranging from the ZAMS to the TAMS \citep{IlievBarzova95,Ilievetal02}.
A majority of \lboo\ stars are pulsating stars in the general class of
\dscuti-type variable stars \citep{Bohlenderetal99}. Such pulsations
indicate that the superficial helium abundance is roughly solar, or higher.
As a much larger fraction of \lboo\ stars than other A-type stars are 
\dscuti\ variables \citep{Paunzenetal98}, it suggests that the \lboo\
phenomenon itself has an effect in their pulsational behavior.
Finally, many \lboo\ stars show signs of circumstellar matter
\citep{Holwegeretal99}.
 
Several ideas have been put forward to account for the \lboo\ phenomenon.
The very peculiar abundances of \lboo\ stars
cannot be reconciled with the diffusion models that have been so
successful for other chemically peculiar A stars such as the FmAm stars
\citep{RMT00}. Nevertheless, as diffusion is an important process in A stars,
it is an important feature of two leading models proposed for
\lboo\ stars, the diffusion/mass loss model
\citep{MichaudCharland86,C93} and the accretion/diffusion model
\citep{VL90,C91}. A third hypothesis calls for a binary 
merging to account for some fraction of \lboo\ stars \citep{Andrievsky97}. 
It also has been shown that a small  number of \lboo\ stars were in 
fact unidentified binaries for which the combined spectra were
mistakenly interpreted as metal-poor \citep{Faraggianaetal01}. 
It is doubtful however that a significant fraction of \lboo\ stars
suffer from this problem.
The pro and cons of these hypothesis have been nicely summarized in 
\citet{Solanoetal01}. It is clear that at this moment no model can
adequately reproduce all the observed properties of \lboo\ stars.

The first challenge facing the models is their inability to
produce both young and old \lboo\ stars.
The diffusion/mass loss
and binary merging models can only yield old \lboo\ stars.
On the other hand the accretion/diffusion model as it stands now
can only occur in young stars before
the circumstellar disk associated with star formation is dissipated. 
 
Currently, the most favored model is the accretion/diffusion model but
new observational data has raised questions about it.
\citet{Heiteretal02} have shown that the composition of \lboo\
stars might not be as consistent with circumstellar gas has thought
previously. They have also raised the question as to why some stars
which have circumstellar disks with ongoing accretion and are similar to \lboo\ stars do
not have peculiar compositions.

This paper concentrates on if and how new insight in the depth of mixing in
slowly rotating F and A type stars \citep{RMT00,RMR01} might affect the accretion model. 
Those models, which will be discussed further in Section~\ref{sec:accdiff}, have shown that
the mixing in slowly rotating stars extends
substantially deeper than the base of the H-He convection zone in A and
B~stars.
The depth of mixing is of crucial importance in
determining the timescales for the formation and persistence of
superficial abundance peculiarities in stars such as the \lboo.
Therefore, as a first step in the computation of more complete models
for \lboo\ stars including as much of the physics of chemical evolution
in stars as possible, the possible effect of deep mixing on timescales
and accretion rates in models of accreting \lboo\ stars will be
discussed here.
A brief overview of the accretion/diffusion model
in the context of deeper mixing will be presented first. Simple models will then be used to
estimate timescales for chemical evolution.

\section{Accretion, Diffusion and Mixing}\label{sec:accdiff}

The model which has met the most success in reproducing
most observed properties of \lboo\ stars is based on the idea of
accretion of dust-depleted matter first suggested by 
\citet{VL90}.  A model featuring accretion and diffusion
was developed by
\citet{C91} (hereafter C91) and a full numerical
simulation in 1 and 2-D in a static model 
was performed by \citet{TC93} (hereafter TC93).
In this scenario, circumstellar matter would be depleted of most heavy
elements as the result of dust formation.
The gas would thereafter
fall back on the star whereas the dust would be pushed out, presumably
by radiation pressure.
While there exists no complete model of the accretion process itself including the
dust-gas separation needed by the accretion scenario,
\citet{AndrievskyPaunzen00} have calculated a simple model for
this which shows that the process is possible under favorable
circumstances.
The accreted matter would form a superficial
layer on the star reflecting the composition of the circumstellar gas.
As the circumstellar matter (supposedly in a disk) would dissipate with
time,
the accretion would stop and diffusion and meridional circulation
would destroy the \lboo\ signature within 1 or 2~Myr. The implications
being that all \lboo\ stars are young and currently undergoing
accretion.
Circumstellar matter has been found around many
but not all \lboo\ stars. This was not thought to be a problem 
as \citet{King94} has argued that the amount of circumstellar gas
needed to permit the formation of abundance anomalies is so small that
the lack of observed circumstellar dust does not necessarily mean that there is
none available for accretion. The results which will follow may
challenge this estimation.

As we proceed we will be assuming that accretion in the spirit of the
idea advanced by \citet{VL90} is a formative
process in at least some \lboo\ stars. Many of the challenges facing the
accretion scenario, such as why some stars with ongoing accretion are
not \lboo\ stars, will not be tackled. The evolution of circumstellar
media is fundamental to the accretion model but is outside the scope of
this paper.
 
As the matter falls on the stellar surface, it is being mixed (diluted)
in a superficial region in which mixing is strong enough to ensure
chemical homogeneity.
The time necessary to establish and destroy the \lboo\ signature
is inversely proportional to the mass of this mixed region and directly
proportional to the flux of
matter in and out of that region. The net flux, in turn, is determined by the
accretion rate and the rate of diffusion (adding mixing, meridional
circulation and any other relevant particle transport process) in
or out of the bottom of the mixed region.

The depth to which superficial mixing extends is a critical parameter in
the evolution of the superficial composition of these stars.
In standard models only the superficial convection zone (SCZ) determined
by the ionization of H and He is fully mixed. There is growing evidence
that there is mixing at greater depths in A stars. 
It has been shown by \citet{RMT00} that the composition of slowly
rotating A stars can only be reproduced by models in which the
superficial mixed zone (SMZ) is much deeper that the SCZ. 
\citet{RMT00} and \citet{RMR01} have both shown that in such slowly
rotating stars, diffusion strongly enhances the composition of iron-peak
elements where they dominate the opacity, which can result in
convection. This convection defines a minimum depth of the mixing well
below the H-He SCZ but it does not explain the even deeper mixing
that is found to be required for the models to fit the observations.  
In more rapidly rotating stars such as the \lboo\ stars, the mixing is
more efficient, as evidenced by the fact that the effect of diffusion is
not seen in rapidly rotating A stars. The mixing found in slowly rotating A stars is expected to be
present as well in fast rotators.
We then expect that mixing in fast rotating A stars will be at least as
deep as in slowly rotating A stars.
Therefore, the mass of the SMZ in \lboo\ stars can be approximated as to be at least as deep as those
of slowly rotating chemically peculiar stars. From 
\citet{RMT00}, the mass of the SMZ is assumed to be greater than $10^{-6}$ times
the stellar mass for the range of stars relevant to \lboo\ stars.
 
Following Eqs~3-5 of C91, one can write the timescale for the evolution 
of the surface composition
for a given accretion rate, a given mass of the SMZ and a given 
chemical species as
\begin{equation}
   \tau = {\rho 4\pi r^2\over M_{\rm SMZ}} \Bigl(v_{\rm
acc}-v\Bigr) \, ,
  \label{eq:tscale}
\end{equation}
as long as $v_{\rm acc}>v$, where all radius dependent variables are
evaluated at the base of the SMZ.
The right hand term describes the net flux of particle at the
base of the SMZ in which the flux created by the accretion ($4\pi\rho
r^2 v_{acc}$) is opposed or enhanced by the flux of matter from
diffusion and other processes ($4\pi\rho r^2 v$).
Each chemical species has a different time scale due to
the specifics of diffusion (i.e. a different $v$) for individual elements.
 
The large scale velocity field due to accretion occurs to preserve hydrostatic
equilibrium.
We assume that the accretion
rate is small enough so the mass of the star can be considered constant
over the timescales which concern us here.
The velocity field is defined by
\begin{equation}
  v_{\rm acc}(r)= {\dot{M}\over 4\pi r^2\rho(r)}
  \label{eq:vacc}
\end{equation}
where $\dot{M}$ is the accretion rate in M$_\sun\,{\rm yr}^{-1}$.
The inverse density dependence of $v_{\rm acc}$ implies that it decreases  
at the base of the SMZ as its depth increases.
 
The transport velocity of particles ($v$ in Eq.~\ref{eq:tscale}) is the sum of the
contributions of diffusion and of advection, such as meridional
circulation.  The diffusion velocity is defined as in
\citet{TRMIR98} where the effects of abundance gradients,
radiation pressure, gravity and temperature gradients are taken into
account. Radiation pressure is included as discussed in
\citet{Richeretal98}. The only advection term included here is
rotationally induced meridional circulation.
The only component of the meridional flow included here is the polar
flow which determines the maximum opposing effect of meridional
circulation to accretion.
A very rough estimate of the polar circulation velocity can be found 
by using Equation~117 of \citet{TassoulTassoul82} with $\mu=1$, and $u(r)$ 
from their Tables~5 through~9 and scaling the
model to the right mass, radius and luminosity.  This has been shown to
be a reasonable approximation by \citet{Charbonneauetal89}.
 
The diffusion velocity depends on the net force exerted on a unit
mass of a specific element which is mainly the difference between gravitation
and radiative pressure. The gravitational part is essentially constant with
depth near the surface but the radiative pressure is very sensitive
to pressure and temperature. The magnitude of the diffusion velocity will generally
decrease with increasing depth but is subject to changing signs as the
radiation pressure dominates the gravity in sections of the star.
The meridional circulation velocity varies less rapidly with depth and its
effect increases relative to diffusion and accretion with increasing
depth. Thus, increasing the depth of the SMZ
will change the flux term of the timescale (Eq~\ref{eq:tscale}),
and the timescales themselves, in a variety of ways. The timescales will
generally increase monotonically and proportionally to the mass of the
SMZ.

\section{Timescales in Models with Deep Mixing}

\begin{figure}
 \plotone{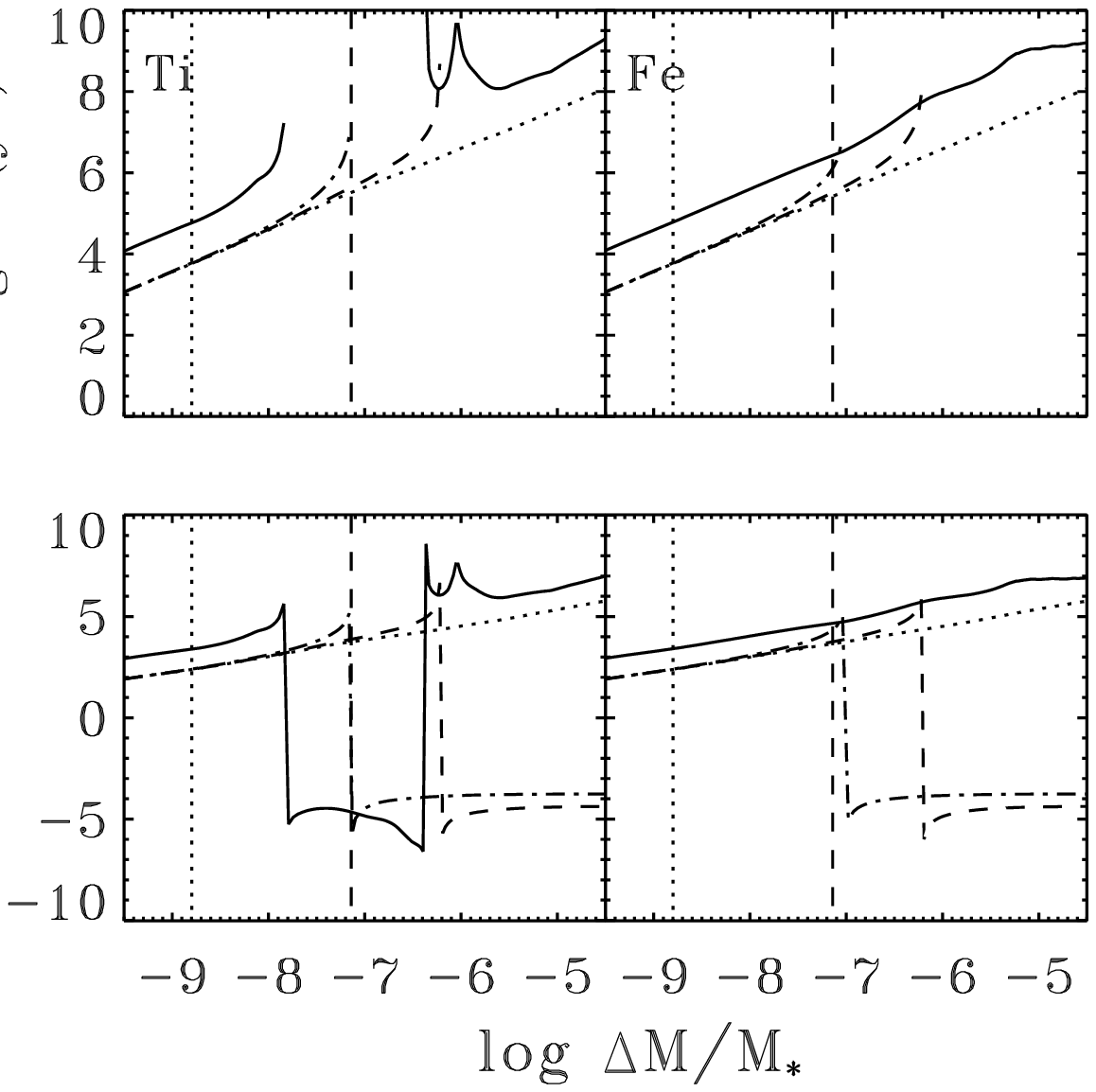}
\caption{The logarithm of the timescale (Eq.~\ref{eq:tscale} using the
velocity $v$ of the lower plots as $v_{acc}-v$) assuming a
SMZ depth equal to the fractional depth $(\Delta M/M_\star)$ (upper plots)
as well as the logarithm of net velocities at the base of the SMZ (lower
plots) are shown for Ti (on the left) and Fe (on the right) as a
function of the fractional depth.  The different curves all refer to the
same cases, the only difference being the microscopic diffusion velocity
of Ti or Fe accordingly. The solid line is for $\dot{M}=10^{-13},
v_{eq}=0.0$; the dotted line for $\dot{M}=10^{-12}, v_{eq}=0.0$;
the dashed line for $\dot{M}=10^{-12}, v_{eq}=50.0$; and the dash-dotted
line for $\dot{M}=10^{-12}, v_{eq}=100.0$ [units being
M$_\sun$\,yr$^{-1}$ and km\,s$^{-1}$ for accretion rate ($\dot{M}$) and equatorial
rotation velocity ($v_{eq}$) respectively]. The net velocity $v$
is always smaller than one, therefore the sign of the velocity in the
plot indicates the direction, negative being toward the surface and
positive being toward the center. The vertical dotted
line indicates the SMZ in standard models of A stars and the vertical
dashed line marks the depth at which $T=200\,000$~K.
\label{fig:fig1}}
\end{figure}

Fig.~\ref{fig:fig1} compares the timescales (Eq~\ref{eq:tscale}) for 
Ti and Fe for a model of a 1.8~M$_\sun$ star at 100~Myr ($T_{eff}=8290$~K, $L=10.7
L_\sun$) with an homogeneous (solar) chemical composition. The Figure
also compares the net velocity with and without meridional circulation velocities.
The diffusion velocities are taken from another model of the same mass
and similar age in which diffusion is computed and in which composition is
not homogeneous (as in \citet{RMT00}). 
Both the timescales and the velocities are shown as a function of depth
in the model to illustrate the effect of increasing the depth of the
SMZ. The standard depth determined by the second ionization of helium
occurs at around $10^{-9}$ in fractional mass whereas the depth of the
SMZ if it reaches the depth of the ``metal opacity bump'' at 200\,00~K
is shown at a little above $10^{-7}$ in fractional mass.
The simple fact of this deep mixing increases the mass of the SMZ by a
factor of at least 50. In fact, \citet{RMT00} show that the range of
mixed mass necessary to account for FmAm stars pushes the mixing to
a typical depth of $-6$ to $-5$. The timescale is then multiplied by a
factor of more than a thousand.

In addition, the radiation pressure for some elements (Ti shown here, and Ca, amongst
others) overwhelms gravity in a large
section of the star centered roughly at the base of the iron convection
zone. This means that there is a flux of matter  opposing the accretion in 
the SMZ. When the meridional circulation is added the upward particle flux 
at the base of the SMZ is even higher, and can dominate accretion even
for elements for which radiation pressure is not significant.
The depth at which meridional circulation dominates over accretion 
depends on the accretion rate and on the rotational velocity of the
star. Diffusion plays only a minor role in sufficiently rapidly rotating
stars.
However, whatever the rotational and diffusion velocities, it is
possible to ensure that the ``right'' abundances are established in the
SMZ in a short enough time by increasing the accretion rate sufficiently.

In all the cases shown in Fig.~\ref{fig:fig1}, 
the net flux in the SMZ is still dominated by accretion because the flux
at the surface is larger than
the flux at the base of the SMZ [$\dot{M} > 4\pi(\rho r^2
(v_{acc}-v))_{SMZ}$], even when diffusion and circulations overwhelm the
accretion flow that the base of the SMZ.
In those cases however, the timescales become too long to expect the
necessary abundances anomalies to form reasonably early in the star's
life.

In the cases of C, N, O, and S, it is generally assumed, as we do here, that their
abundance is normal in the accreted matter. As these elements are not
significantly supported by radiative pressure, their abundance will remain
normal in \lboo\ stars provided that the accretion rate is large enough
to compensate the gravitational settling at the base of the SMZ. This is 
easily satisfied in the models and so they do not provide significant
constraints on the accretion rates and particle transport processes.

\section{Discussion}

It has been shown that indications of deep mixing in F, A and B stars
\citep{RMT00}
can have a significant effect on the accretion/diffusion model
for \lboo\ stars, leading to larger predicted accretion rates 
or longer timescales for the formation of the requisite surface
composition.
This assumes that the mixing in slowly rotating stars
is similarly active in faster rotators.
Such an extrapolation, as is done here, is still founded on
circumstantial evidence and is subject to confirmation.

Nevertheless, deep mixing in \lboo\ stars raises intriguing possibilities
regarding the few but important points of contention between the
standard accretion/diffusion model \citep{TC93} and the observations
\citep{Solanoetal01,Heiteretal02}.
One of the most difficult problem facing the accretion model is the 
existence of old \lboo\ stars. In A type stars, circumstellar disks
are not expected to persist more than a couple of hundred Myrs
\citep{MeyerBeckwith00} which is far less than the oldest \lboo. 
If one assumes that \lboo\ stars are mixed to a depth of $10^{-6}$
in fractional mass, as argued here, and that the accretion rate was high enough early
on to ensure the observed abundances reflect those of the accreted matter,
then the larger timescale for the evolution of the surface
abundance might provide a way to explain older \lboo\ stars.

With a standard SMZ only a deep as the SCZ, the time needed to erase the
\lboo\ signature is of the order of 1~Myr \citep{TC93}.
The timescale in the case of deep mixing will be increased by a factor
of some few hundreds because of the increase in the mass of the SMZ, but
this is mitigated by the relative increase in flux due to meridional circulation.
The net effect for a mixed mass of $10^{-6}$~M$_\star$ is an increase of the timescale from
by a factor of 5 only. 

\begin{figure}
\plotone{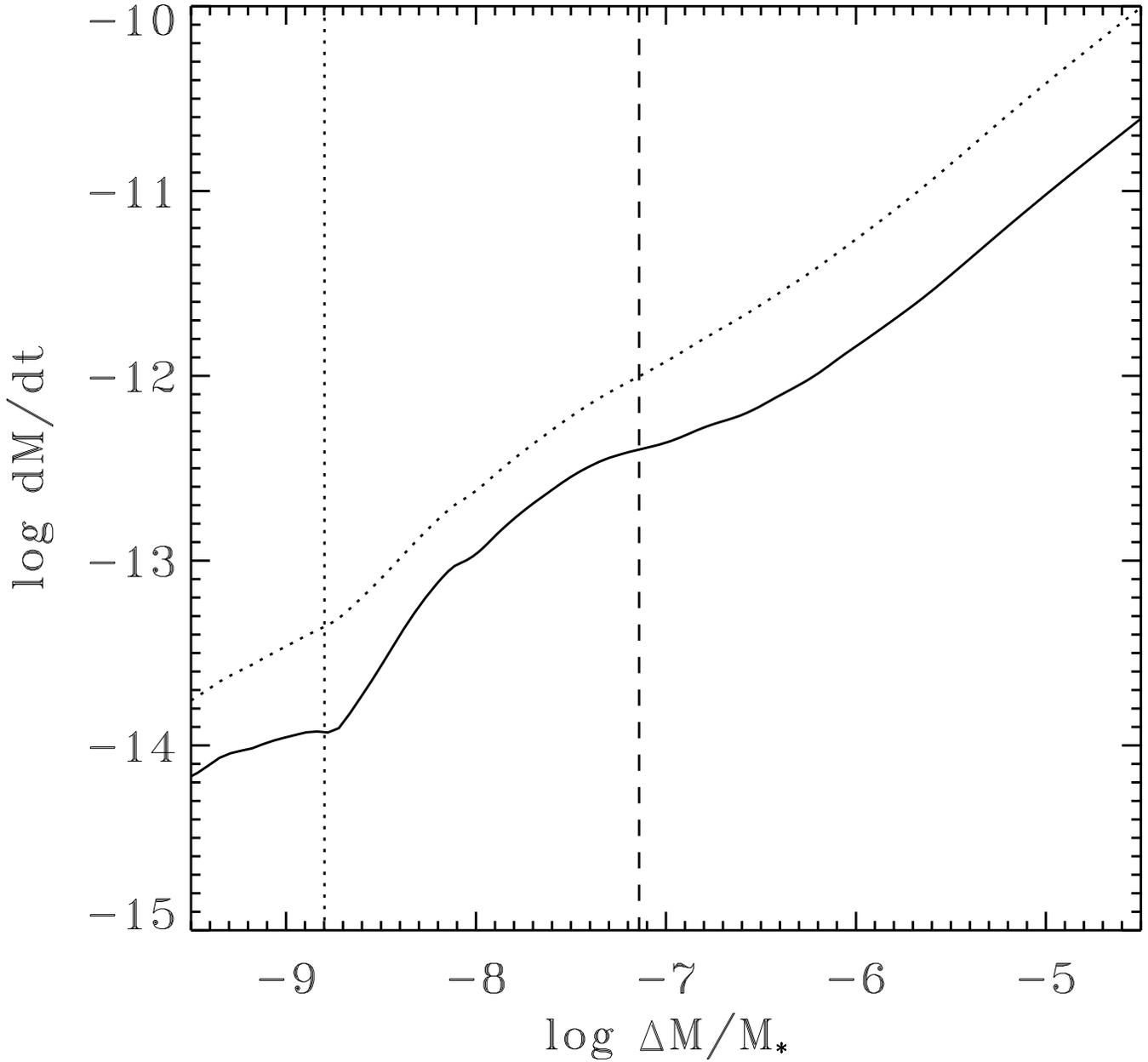}
\vskip 0.5cm
\caption{The minimum accretion rate needed to balance the diffusion $+$ meridional circulation velocity
         of Ti at the base of the SMZ is shown for an equatorial rotation
         velocity of 50 (solid line) and 100~km$\,s^{-1}$ (dotted line). Vertical lines are as in 
         Fig.~\ref{fig:fig1}.
         \label{fig:mloss}}
\end{figure}

A complicating factor for the accretion scenario is that not only is it
necessary to dramatically increase the amount of gas accreted on the
star in order to impart the composition of dust-depleted circumstellar
gas to the SMZ, which may be accounted by much larger accretion rates
on the pre-main-sequence, but much larger ongoing accretion rates are necessary 
to sustain the abundance peculiarities if the SMZ is as deep as
suggested here. Fig.~\ref{fig:mloss}, shows that an accretion rate of
$10^{-12}$ to $10^{-11}$~M$_\sun\,yr^{-1}$ is necessary to just balance the 
flux of particles entering or leaving the SMZ at its base from diffusion
and meridional circulation. Such large rates may not problematic as they
have been claimed in $\beta$~Pictoris \citep{Beustetal96}. It might however
raise questions as to whether the amount of circumstellar matter
required to provide such large rates could remain unseen as is the case
in many \lboo\ stars.
Still, if one assumes that it is the case and that the necessary
accretion is ongoing as long as the circumstellar disk is present,
one would still not expect \lboo\ stars as old as 1~Gyr. 

We have restricted ourselves to rotational velocities of 
100~km\,s$^{-1}$ or lower because of the limitations of the formalism
for meridional circulation used here. \lboo\ stars can rotate at a much
faster rate, as much as 250~km\,s$^{-1}$ \citep{Paunzen01}. 
In such stars, the meridional circulation would dominate
a given accretion rate for much shallower SMZs. The accretion rates
required to establish the \lboo\ signature could then be an order of
magnitude larger, or more, than those found for the models discussed here.

Finally, as \lboo\ stars often are pulsating stars, it is tantalizing to imagine that
there might be a seismic signature of the depth of the mixing
considering that the abundance of most metals is completely different in 
the ``metal opacity bump'' depending on the models discussed here.
As the metals play a role in driving pulsations and determining the
structure of the envelope in these stars, accretion with deep mixing 
might yield an observable signature in either which modes become
overstable or in shifts in the frequencies of modes of pulsations
with respect to standard models for \dscuti\ stars. Preliminary 
models have shown shifts in frequencies by as much as 10 to 30\% \citep{T00}
but a seismic test for mixing in \lboo\ stars, only possible with
reliable mode identification, remains out of our reach 
at this point in time.

\lboo\ stars are perhaps the best candidates to provide constraints on
mixing mechanisms in moderately rapidly rotating early type stars for
which more standard diagnostics such as lithium abundances are not 
available. The major observational effort spent on these stars in
recent years and their inclusion as main targets for planned
asteroseismology experiments make a parallel theoretical effort
necessary. 
Only more sophisticated calculations of evolving A stars with accretion
and a more precise treatment of rotational mixing and circulation will
determine if the accretion/diffusion model can be reconciled with the
challenges that the recent observational studies of these stars have uncovered.

\acknowledgments

I gratefully acknowledge the comments of Dr. Georges Michaud and an
anonymous referee from which this paper has benefited greatly.
This work was performed under the auspices of the U.S.
Department of Energy, National Nuclear Security Administration by the
University of California, Lawrence Livermore National Laboratory under
contract No.W-7405-Eng-48.

\clearpage


\clearpage 

\clearpage 

\end{document}